\documentclass[a4paper,10pt]{article}
\usepackage[margin=1.0in]{geometry}

\usepackage{amssymb,amsfonts,amsmath,epsfig,float,graphicx}

\usepackage {url}

\newcommand{\qed}{\nobreak \ifvmode \relax \else
      \ifdim\lastskip<1.5em \hskip-\lastskip
      \hskip1.5em plus0em minus0.5em \fi \nobreak
      \vrule height0.75em width0.5em depth0.25em\fi}

\def\upd{{\rm d}}
\newcommand{\be}{\begin{eqnarray}}
\newcommand{\ee}{\end{eqnarray}}
\newcommand{\Tr}{\,{\rm Tr}\,}

\def\Io{{\mathbb I}}

\def\Ro{{\mathbb R}}

\def\kk{{\bf k}}

\def\xx{{\bf x}}

\def\uc{q}
\def\Ecl{E_{\mbox{\tiny cl}}}
\def\lu{{l}}

\renewcommand{\Re}{\,{\rm Re}\,}

\begin{document}
\title{Reducible Quantum Electrodynamics.\\
II. The charged states of the vacuum}
\author{Jan Naudts}

\maketitle

\begin{abstract}
An explicit construction is given of field operators satisfying the free Dirac equation.
The quantum expectation of these field operators forms a spinor which satisfies
the original Dirac equation. The current operators are defined as pair correlation
functions. Explicit expressions in terms of creation and annihilation operators
are obtained. A small example shows that all relevant quantities are
mathematically well-defined.
\end{abstract}

\section*{Introduction}

\def\mstar{\kappa}

In the first of this series of papers \cite {NJ15a} free electromagnetic fields
are treated in a reducible representation. The methodology developed there
is used here to present a theory of free fermionic fields.

A characteristic feature of fermionic fields are their anti-commutation relations.
For scalar fields $\hat\phi(x)$ they usually read
\be
\{\hat\phi(x),\hat\phi(y)\}_+=0
\quad\mbox{ and }\quad
\{\hat\phi(x),\hat\phi^\dagger(y)\}_+=\Delta(x-y)
\label{intro:car}
\ee
with the Pauli-Jordan function defined by
\be
\Delta(x)&=&i\int\upd\kk\,\frac{1}{(2\pi)^32\omega(\kk)}\left(e^{ik_\nu x^\nu}-e^{-ik_\nu x^\nu}\right).
\ee
Here, the convention $k_0=\omega(\kk)/c$ is used, with the dispersion relation given by
\be
\omega(\kk)&=&\sqrt{\mstar^2+|\kk|^2}
\ee
and with $\mstar=mc/\hbar$ the mass of the fermion expressed in appropriate units.

In the present approach the anti-commutation relations are non-canonical.
Typical for the approach of \cite {NJ15a}
is that the integrations over wave vectors move from operator
expressions towards the evaluation of expectation values.
The  equivalent of (\ref {intro:car}) becomes
\be
\{\hat\phi_\kk(x),\hat\phi_{\kk'}(x')\}_+=0
\quad\mbox{ and }\quad
\{\hat\phi_\kk(x),\hat\phi^\dagger_{\kk'}(x')\}_+=e^{ik_\nu x^\nu}e^{-ik'_\nu x'^\nu}.
\label{intro:car2}
\ee
See the next section.
The generalization of (\ref {intro:car2})
to the electron field is found in Section \ref {sect:electron} and is done in an obvious way.

These the anti-commutation relations differ in an essential way from the 
non-canonical commutation relations introduced in \cite {CM04}.
In the present notations they read
\be
\{\hat\phi_\kk,\hat\phi_{\kk'}\}_+=0
\quad\mbox{ and }\quad
\{\hat\phi_\kk,\hat\phi^\dagger_{\kk'}\}_+=(2\pi)^3 2k_0\delta(\kk-\kk').
\ee

Given creation and annihilation operators for the electron field
one can introduce new operators satisfying the Dirac equation.
This is done in Section \ref {sect:dirac}.
The main advantage of using Dirac's equation
is the availability of an expression for the electric current in terms of Dirac fields.
The present paper defines the electric current as a pair correlation
function of Dirac fields. A short calculation then shows that the definition
coincides with the well-known definition of the Dirac current.

In Part I it is mentioned that free electromagnetic fields can be seen as living in two additional
dimensions, this is, in a world of 4+2 dimensions. The origin for this view is the
picture of the e.m.~field as a collection of two-dimensional harmonic oscillators, one in each point
of space. For the electron field no extra dimensions are needed. It suffices to add one spin variable and one
charge variable for each wave vector $\kk$. One can see these two variables as a
local description of the properties of space. The electromagnetic waves propel through
a two-dimensional space which is locally deformed by spin and charge.
This explains the title of the present paper.

\section{Fermionic fields}

\subsection{The Klein-Gordon equation}

This section concerns the quantum field description of fermions with
a rest mass $m$. The appropriate wave equation is
the Klein-Gordon equation
\be
(\square +\mstar^2)\phi(x)=0
\quad\mbox{with }\mstar=\frac{mc}{\hbar}.
\label{fermion:kg}
\ee
For $m=0$ it reduces to the d'Alembert equation $\square\phi=0$,
discussed in Part I.

Propagating wave solutions are of the same form as in Part I
\be
\phi(x)
&=&
2\Re \int_{\Ro^3}\upd\kk \,\frac{\lu}{N(\kk)}
f(\kk)e^{-ik_\mu x^\mu},
\label{weq:exp}
\ee
but with a dispersion relation given by the positive square root
\be
k^0\equiv\omega(\kk)/c=\sqrt{\mstar^2+|\kk|^2},
\ee
and a corresponding normalization
\be
N(\kk)=\sqrt{(2\pi)^32k_0}.
\ee
The constant $\lu$ is inserted in (\ref {weq:exp})
for dimensional reasons. 
It makes $|f(\kk)|^2$ into a density.
Its value is discussed later on. 

\subsection{Larmor precession}
\label{fermion:larmor}

We use the harmonic oscillator in the description of bosons because it
exhibits periodic motion.
An alternative model exhibiting periodic motion is that of Larmor precession.
It involves the Pauli matrices $\hat\sigma_\alpha$, $\alpha=1,2,3$.
The time evolution is
\be
\hat \sigma_1(t)&=&\hat \sigma_1(0)\cos(\omega t)+\hat \sigma_2(0)\sin(\omega t),\\
\hat \sigma_2(t)&=&\hat \sigma_2(0)\cos(\omega t)-\hat \sigma_1(0)\sin(\omega t),\\
\hat \sigma_3(t)&=&\hat \sigma_3(0).
\ee
The Hamiltonian reads
\be
\hat H=-\frac{1}{2}\hbar\omega \hat \sigma_3.
\label{fermion:larmorham}
\ee
Note that
\be 
\hat \sigma_\pm(t)=\hat \sigma_\pm(0)e^{\mp i\omega t},
\ee
where 
\be
\hat \sigma_\pm=\frac 12(\hat \sigma_1\pm i\hat \sigma_2).
\ee

A wave function $|z\rangle$ describing the state of the system consists of two complex numbers
$z_+$, $z_-$ which satisfy the normalization condition
$|z_+|^2+|z_-|^2=1$.
The quantum expectation of the Hamiltonian (\ref {fermion:larmorham}) equals
\be
\langle z|\hat H |z\rangle &=&-\frac{1}{2}\hbar\omega(\kk)
(\overline{z_1}\quad\overline{z_2})
\left(\begin{array}{lr}
       1 &0\\ 0 &-1
      \end{array}\right)
\left(\begin{array}{c}
       z_1\\z_2
      \end{array}\right)\cr
&=&-\frac 12\hbar\omega(|z_1|^2-|z_2|^2).
\ee
The quantum expectation of the matrices $\hat \sigma_\pm(t)$ is given by
\be
\langle z|\hat \sigma_+(t)|z\rangle =\frac 12\overline{z_1}z_2e^{-i\omega t}
\quad\mbox{and}\quad
\langle z|\hat \sigma_-(t) |z\rangle =\frac 12\overline{z_2}z_1e^{i\omega t}.
\ee

\subsection{Field operator}

We now proceed by analogy with the bosonic case, in particular
the case of the single photon states.
The Hamiltonian (\ref{fermion:larmorham}) becomes
\be
\hat H=\frac{1}{2}\hbar\omega(\kk) (\Io-\hat \sigma_3).
\label{fermion:hamplus}
\ee
A constant matrix has been added to make the Hamiltonian non-negative.
This does not change the dynamics of the Larmor precession.
The frequency $\omega$ is now a $\kk$-dependent multiplication operator.
Also the wave function $|z\rangle$ becomes $\kk$-dependent.
We write, similar to the case of single photon states,
\be
\psi_\kk=\left(\begin{array}{c}
               \sqrt{1-\rho(\kk)}e^{i\chi(\kk)}\\
	       \sqrt{\rho(\kk)}e^{i\xi(\kk)}
               \end{array}\right).
\label{fermion:wf}
\ee
Then the quantum expectation of the Hamiltonian becomes
\be
E
&=&\lu^3\int\upd\kk\,\langle\psi_\kk|\hat H\psi_\kk\rangle\cr
&=&\lu^3\int\upd\kk\,\hbar\omega(\kk)\rho(\kk).
\label{fermion:quantenerg}
\ee
This reveals that $\rho(\kk)$ is a distribution of wave vectors.
It is restricted by the condition that $0\le\rho(\kk)\le 1$ for all $\kk$.
Because the energy must remain finite the distribution $\rho(\kk)$ should go to 0
fast enough for large values of the frequency $\omega(\kk)=c|\kk|$.

Introduce now the field operator
\be
\hat\phi(x)=\hat\sigma_+(t)e^{i\kk\cdot\xx}+\hat\sigma_-(t)e^{- i\kk\cdot\xx}.
\label{fermion:fieldop}
\ee
It is tradition to decompose this field operator into so-called positive-frequency and negative-frequency
operators
\be
\hat\phi(x)&=&\hat\phi^{(+)}(x)+\hat\phi^{(-)}(x), \quad\mbox{ with }\cr
\hat\phi^{(+)}(x)&=&\hat\sigma_+(t)e^{i\kk\cdot\xx}=\hat\sigma_+(0)e^{-ik_\nu x^\nu},\cr
\hat\phi^{(-)}(x)&=&\hat\sigma_-(t)e^{- i\kk\cdot\xx}=\hat\sigma_-(0)e^{ik_\nu x^\nu}
=\left(\hat\phi^{(+)}(x)\right)^\dagger.
\ee
They satisfy the anti-commutation relations
\be
\hat\phi^{(+)}(x)\hat\phi^{(+)}(y)=0
\quad\mbox{ and }\quad
\{\hat\phi^{(+)}(x),\hat\phi^{(-)}(y)\}_+=e^{-ik_\mu (x-y)^\mu}.
\label{fermion:car}
\ee
The r.h.s.~of the latter is still a multiplication operator. Therefore,
these anti-commutation relations are non-canonical.

The classical field $\phi(x)$ corresponding with the field operator $\hat\phi(x)$ equals
\be
\phi(x)
&=&2\Re \int\upd\kk \,\frac{\lu^{5/2}c^{1/2}}{\sqrt{(2\pi)^32\omega(\kk)}}
\sqrt{\rho(\kk)(1-\rho(\kk))}e^{-i(\chi(\kk)-\xi(\kk))}e^{-ik_\mu x^\mu}.
\label{fermion:classfield}
\ee
The integral converges provided that the density $\rho(\kk)$ tends fast enough
 either to 0 or to 1 for large values of $|\kk|$.
The expression (\ref {fermion:classfield}) is
of the form (\ref {weq:exp}) with
\be
f(\kk)&=&\lu^{3/2}\sqrt{\rho(\kk)(1-\rho(\kk))}e^{-i(\chi(\kk)-\xi(\kk))}.
\ee

Hence, the energy of the classical field $\phi(x)$ is given by (see Part I)
\be
\Ecl
&=&\int\upd\kk\,\hbar\omega(\kk)|f(\kk)|^2\cr
&=&\lu^3\int\upd\kk\,\hbar\omega(\kk)\rho(\kk)(1-\rho(\kk)).
\label{fermion:classenerg}
\ee 
For small values of $\rho(\kk)$ the quantum mechanical and the classical
energies coincide. For large values the quantum energy grows linearly while
the classical value vanishes again when $\rho(\kk)$ tends to 1. 
One concludes that always part of the energy of a fermion is of a quantum nature.
This contrasts with the situation for photons. Their quantum contribution to the energy
vanishes in the case of coherent states. On the other hand, their classical
energy vanishes for photonic eigenstates.


\section{The electron}
\label{sect:electron}

\subsection{The algebra of creation and annihilation operators}

The electron wave is fermionic. It has also two polarizations,
which are related to the spin of the electron.
In addition, it has an anti-particle, which is the positron.
This means that the electron field has 2 internal degrees of freedom
and that we need 4 copies of the spin matrices $\hat \sigma_\pm$
instead of the single copy introduced in the Section \ref {fermion:larmor}.
The corresponding field operators are denoted $\hat\phi_s(x)$, with the index $s$
running from 1 to 4.
Each of them satisfies the anti-commutation relations (\ref {fermion:car}).
Together they generate an algebra known as the Clifford algebra.
An explicit representation of the operators as 16-by-16
matrices is easily constructed (see for instance Section 3-9 of \cite{JR76}).
However, it is not needed in the sequel. All we need is that each individual operator
$\hat\phi_{s,\kk}(x)$ can be written in the form (\ref {fermion:fieldop})
and that together they satisfy the anti-commutation relations 
\be
\{\hat\phi_{s,\kk}^{(+)}(x),\hat\phi_{t,\kk'}^{(+)}(y)\}_+=0
\quad\mbox{ and }\quad
\{\hat\phi_{s,\kk}^{(+)}(x),\hat\phi_{t,\kk'}^{(-)}(y)\}_+=
\delta_{s,t}e^{-ik_\mu x^\mu}e^{ik'_\mu y^\mu}.
\label{electron:car}
\ee
Note that (\ref {fermion:fieldop}) implies that
\be
\hat\phi_{s,\kk}^{(+)}(x)&=&e^{-ik_\mu x^\mu}\hat\phi_{s,\kk}^{(+)}(0).
\ee

A familiar notation for these operators, evaluated at $x=0$, is
\be
b_{\uparrow}=\hat\phi^{(+)}_{1}(0),\quad
b_{\downarrow}=\hat\phi^{(+)}_{2}(0),\quad
d_{\downarrow}=\hat\phi^{(+)}_{3}(0),\quad
d_{\uparrow}=\hat\phi^{(+)}_{4}(0).
\ee
This alternative notation is not used here.

An arbitrary basis vector of the 16-dimensional Hilbert space at constant wave vector $\kk$
is specified by a subset $\Lambda\subset\{1,2,3,4\}$
and is given by
\be
|\Lambda\rangle&=&
[\hat\phi_{4,0}^{(-)}]^{\Io_{4\in\Lambda}}
[\hat\phi_{3,0}^{(-)}]^{\Io_{3\in\Lambda}}
[\hat\phi_{2,0}^{(-)}]^{\Io_{2\in\Lambda}}
[\hat\phi_{1,0}^{(-)}]^{\Io_{1\in\Lambda}}|\emptyset\rangle.
\ee
For instance, if $\Lambda=\{1,3\}$ then
$|\{1,3\}\rangle=\hat\phi_{3,0}^{(-)}\hat\phi_{1,0}^{(-)}|\emptyset\rangle$.

\subsection{The Hamiltonian}

The Hamiltonian of the electron field is the sum of 4 copies of the scalar
Hamiltonian (\ref {fermion:hamplus}).
Note that
\be
\hat\phi^{(-)}_{s}\hat\phi^{(+)}_{s}
&=&\hat\sigma_-(0)\hat\sigma_+(0)\cr
&=&\frac{1}{2}\left(1-\hat\sigma_3(0)\right).
\ee
Hence the Hamiltonian can be written as
\be
\hat H_\kk&=&\hbar\omega(\kk)\sum_{s=1}^4\hat\phi^{(-)}_{s}\hat\phi^{(+)}_{s}.
\ee
Note that the Hamiltonian is positive.
It is tradition to assign negative energies to positrons and positive energies to electrons.
This tradition is not followed here because it does not make sense.
It is a remainder of Dirac's interpretation of positrons as holes in a sea of electrons.
The alternative path assigns the vacuum state to one of the eigenstates of $\hat\sigma_3(0)$
instead of assigning a particle/anti-particle pair to the two eigenstates.
The dimension of the Hilbert space goes up from 4 (the number of components of a Dirac spinor)
to 16. This is meaningful because the Dirac equation considered here is an equation for
field operators and not the original one which holds for classical field spinors
(see (\ref {electron:diracclass}) below).

Number operators $\hat N^{(s)}$ are defined by
\be
\hat N^{(s)} &=&\hat\phi^{(-)}_{s}(0)\hat\phi^{(+)}_{s}(0)
\quad s=1,2,3,4.
\label{electron:numop}
\ee
They do not depend on the wave vector $\kk$.
They appear in the Hamiltonian
\be
\hat H_\kk&=&\hbar\omega(\kk)\sum_{s=1}^4\hat N^{(s)}.
\label{electron:ham}
\ee

The field operators $\hat\phi_{s,\kk}^{(+)}(x)$ satisfy Heisenberg's equations of motion
\be
i\hbar c\frac{\partial\,}{\partial x^0}\hat\phi_{s,\kk}^{(+)}(x)
&=&\left[\hat\phi_{s,\kk}^{(+)}(x), \hat H_\kk\right]_-.
\ee

\subsection{Dirac's equation}
\label{sect:dirac}

Introduce the gamma matrices. In the standard representation they read
\be
\gamma_0=\left(\begin{array}{lr}
                \Io &0\\0 &-\Io
               \end{array}\right)
\quad\mbox{and}\quad
\gamma_\alpha=\left(\begin{array}{lr}
                0 &-\sigma_\alpha\\\sigma_\alpha &0
               \end{array}\right).
\ee
Next introduce auxiliary field operators $\hat\psi_r$, $r=1,2,3,4$, defined by
\be
\hat\psi_{r,\kk}(x)
&=&
\sum_{s=1,2}u_r^{(s)}(\kk)\hat\phi_{s,\kk}^{(+)}(x)
+\sum_{s=3,4}v_r^{(s)}(\kk)\hat\phi_{s,\kk}^{(-)}(x).
\label{electron:defpsi}
\ee
The vectors $u^{(1)},u^{(2)},v^{(3)},v^{(4)}$ are the analogues of the polarization vectors of the photon.
They are partly fixed by the requirement that the vector
with components $\hat\psi_r$ satisfies Dirac's equation
\be
i\gamma^\mu\partial_\mu \hat\psi(x)=\mstar \hat\psi(x).
\label{electron:dirac}
\ee
Indeed, using
\be
\partial_\mu \hat\phi_{s,\kk}^{(\pm)}&=&\mp ik_\mu \hat\phi_{s,\kk}^{(\pm)}
\ee
one finds that a sufficient condition for (\ref {electron:dirac}) to hold is
\be
\gamma^\mu k_\mu u^{(s)}=\mstar u^{(s)}
\quad\mbox{and}\quad
\gamma^\mu k_\mu v^{(s)}=-\mstar v^{(s)}.
\ee
Each of these two equations has two independent solutions.
See the Appendix \ref{appendix:polelectron:section}.
They can be chosen to satisfy the orthogonality relations
\be
\sum_r\overline{u^{(s)}_r}(\kk)u^{(s')}_r(\kk)&=&\delta_{s,s'},\cr
\sum_r\overline{v^{(s)}_r}(\kk)v^{(s')}_r(\kk)&=&\delta_{s,s'},\cr
\sum_r\overline{u^{(s)}_r}(\kk)v^{(s')}_r(-\kk)&=&0.
\label {electron:ortho}
\ee

For further use note the inverse relations
\be
\sum_r\overline{u_r^{(s)}(-\kk)}\hat\psi_{r,\kk}(x)
&=&\frac{\mstar}{k_0}\hat\phi^{(+)}_{s,\kk}(x),
\qquad s=1,2,
\label{electron:inv+}\\
\sum_r\overline{v_r^{(s)}(-\kk)}\hat\psi_{r,\kk}(x)
&=&\frac{\mstar}{k_0}\hat\phi^{(-)}_{s,\kk}(x),
\qquad s=3,4.
\label{electron:inv-}
\ee

Some further properties are (see the Appendix \ref {app:useful})
\be
\langle u^{(s)}(\kk)|\gamma^0\gamma^\mu u^{(s')}(\kk)\rangle
&=&
\frac{k^\mu}{k_0}\delta_{s,s'},
\quad s,s'=1,2,
\label{electron:uid}
\ee
and
\be
\langle v^{(s)}(\kk)|\gamma^0\gamma^\mu v^{(s')}(\kk)\rangle
&=&
\frac{k^\mu}{k_0}\delta_{s,s'},
\quad s,s'=3,4.
\label{electron:vid}
\ee

It is easy to see, using the anti-commutation relations, that
\be
\left\{\hat\psi_{r,\kk}(x),\hat\psi_{r',\kk'}(y)\right\}_+
&=&0
\ee
holds for any choice of parameters and indices.
On the other hand,
\be
\left\{\hat\psi_{r,\kk}(x),\hat\psi^\dagger_{r',\kk}(x)\right\}_+
&=&\delta_{r,r'}
\ee
holds only for equal positions and momenta. More general is
\be
\left\{\hat\psi_{r,\kk}(x),\hat\psi^\dagger_{r',\kk'}(y)\right\}_+
&=&\sum_{s=1,2}\overline{u^{(s)}_{r'}(\kk')}u^{(s)}_{r}(\kk)e^{-i(k_\nu x^\nu-k'_\nu y^\nu)}\cr
& &
+\sum_{s=3,4}\overline{v^{(s)}_{r'}(\kk')}v^{(s)}_{r}(\kk)e^{i(k_\nu x^\nu-k'_\nu y^\nu)}.
\ee

Given wave functions $\psi_\kk$ of the form
\be
\psi_\kk&=&\sum_{\Lambda\subset\{1,2,3,4\}\}}z^{\Lambda}_{\kk}|\Lambda\rangle,
\ee
with complex coefficients $z^{\Lambda}_{\kk}$ satisfying
\be
\sum_{\Lambda\subset\{1,2,3,4\}}|z^{\Lambda}_{\kk}|^2&=&1,
\qquad\mbox{ for all }\kk,
\ee
a Dirac spinor containing classical fields is defined by
\be
\phi_r(x)
&=&\int\upd\kk\,\frac{\lu^{5/2}c^{1/2}}{\sqrt{(2\pi)^32\omega(\kk)}}\langle\psi_\kk|\hat\psi_{r,\kk}\psi_\kk\rangle.
\label{electron:classfield}
\ee
It satisfies the Dirac equation
\be
i\gamma^\mu\partial_\mu \phi(x)=\mstar\phi(x).
\label{electron:diracclass}
\ee

\subsection{The adjoint equation}

\def\psicha{\hat\psi^{\mbox{\tiny a}}}
\def\psichap{\hat\psi^{\mbox{\tiny a}\prime}}

The so-called adjoint spinor is defined by
\be
\psicha_r(x)&=&\sum_{r'}\hat\psi^\dagger_{r'}(x)\gamma^0_{r',r}.
\ee
It satisfies the adjoint equation
\be
-i\partial_\mu \sum_r\psicha_r(x)\gamma^\mu_{r,r'}&=&\mstar \psicha_{r'}(x).
\label{fermion:diracadjoint}
\ee
To prove this take the adjoint of the Dirac equation and multiply with $\gamma^0$ from the right.
This gives
\be
-i\partial_\mu\sum_{r,r'}\hat\psi^\dagger_{r'}(x)(\gamma^\mu)^\dagger_{r',r}\gamma^0_{r,r''}
&=&\mstar\psicha_{r''}(x).
\ee
Next use that $(\gamma^\mu)^\dagger\gamma^0=\gamma^0\gamma^\mu$ to obtain (\ref {fermion:diracadjoint}).

The charge conjugation matrix $C$ is defined by 
\be
C\gamma^\mu C^{-1}=-(\gamma^\mu)^{\rm T}.
\ee
Using the standard representation of the gamma matrices it equals
$C=i\gamma^2\gamma^0$. See for instance Section 10.3.2 of \cite{GR96}.
Its main properties are
\begin{itemize}
 \item $C^{-1}=C^\dagger=C^{\rm T}=-C$;
 \item $\sum_{r'}C_{r,r'}\overline {u_{r'}^{(1)}(\kk)}=v^{(4)}_r(-\kk)$;
 \item $\sum_{r'}C_{r,r'}\overline {u_{r'}^{(2)}(\kk)}=v^{(3)}_r(-\kk)$.
\end{itemize}
The charge conjugation operator $\hat C$ is defined by $\hat C^\dagger=\hat C^{-1}$ and
\be
\hat C\hat \psi_r(x)\hat C^{-1}&=&\sum_{r'}C_{r,r'}\psicha_{r'}(x).
\ee
One verifies that
\be
\hat C \psicha_{r}(x)\hat C^{-1}&=&-\sum_{r'}C_{r,r'}\hat\psi_{r'}(x).
\ee

Charge conjugation is a symmetry of the Dirac equation.
More precisely, it maps the Dirac equation onto the adjoint equation.
Indeed, from (\ref {electron:dirac}) follows
\be
\hat C\left(
\sum_{r'}\gamma^\mu_{r,r'} i\partial_\mu \hat \psi_{r'}(x)-\mstar \hat\psi_r(x)\right)
\hat C^{-1}
&=&
\sum_{r'}\gamma^\mu_{r,r'}i\partial_\mu\sum_{r''}C_{r',r''}\psicha_{r''}(x)
-\mstar \sum_{r'}C_{r,r'}\psicha_{r'}(x)\cr
&=&\sum_{r'}C_{r,r'}\left(-\sum_{r''} i\partial_\mu\psicha_{r''}(x)\gamma^\mu_{r'',r'}
-\mstar\psicha_{r'}(x)\right).
\ee
This means that $\hat\psi(x)$ is a solution of the Dirac equation
if and only if $\psicha(x)$ is a solution of the adjoint equation.

\subsection{Two-point correlations}

\def\pc{r}
\def\Pc{R}

\def\Ga{G^{\mbox{\tiny a}}}

A two-point correlation function for the Dirac field operators $\hat\psi_r(x)$ 
is defined by (compare with that introduced in Part I of the paper)
\be
\Ga_{r',r}(x,x')
&=&
\int\upd\kk \,\frac{\lu^{5/2}c^{1/2}}{\sqrt{(2\pi)^32\omega(\kk)}}
\int\upd\kk' \,\frac{\lu^{5/2}c^{1/2}}{\sqrt{(2\pi)^32\omega(\kk')}}
\langle\psi_{\kk}|\psicha_{r,\kk}(x)\hat\psi_{r',\kk'}(x')\psi_{\kk'}\rangle.\cr
& &
\label{fermion:twopointmod}
\ee
Note the order of the indices $r,r'$.
By use of Dirac's equation there follows
\be
i\mstar \Ga_{r',r}(x,x')
&=&-\frac{\partial\,}{\partial {x'}^{\mu}}\sum_{r''}\gamma^\mu_{r',r''}\Ga_{r'',r}(x,x').
\label{fermion:twopointeq1}
\ee
On the other hand, using the adjoint equation, one obtains
\be
i\mstar \Ga_{r',r}(x,x')
&=&\frac{\partial\,}{\partial {x}^\mu}\sum_{r''}\Ga_{r',r''}(x,x')\gamma^\mu_{r'',r}.
\label{fermion:twopointeq2}
\ee
Subtracting one expression from the other yields
\be
0&=&\frac{\partial\,}{\partial {x'}^\mu}\sum_{r''}\gamma^\mu_{r',r''}\Ga_{r'',r}(x,x')
+\frac{\partial\,}{\partial {x}^\mu}\sum_{r''}\Ga_{r',r''}(x,x')\gamma^\mu_{r'',r}.
\ee
Now take $r=r'$ and sum over $r$. There follows
\be
0&=&\frac{\partial\,}{\partial {x'}^\mu}\Tr \gamma^\mu \Ga(x,x')
+\frac{\partial\,}{\partial {x}^\mu}\Tr \Ga(x,x')\gamma^\mu.
\ee
In particular, one has
\be
0&=&\frac{\partial\,}{\partial {x}^\mu}\Tr \gamma^\mu \Ga(x,x).
\label{fermion:cont}
\ee
This result shows that the vector $\pc(x)$  with 4 components
\be
\pc^\mu(x)=\Tr \gamma^\mu \Ga(x,x)
\ee
satisfies the continuity equation.

\subsection{Properties of the $\pc$-current}

The vector $\pc (x)$, introduced above, describes a current,
which however is not yet the electric current.
The latter is introduced in the next section.
It is tempting to interpret $\pc (x)$ as the particle current.
However, this interpretation has some difficulties.
The integration over space of its zeroth component,
which should be the total number of particles,
is usually divergent. The latter is anyhow not a very
interesting quantity once the interaction with
the electromagnetic field is turned on because it is not conserved.
An electron and a positron may annihilate each other
or may be created by a pair of photons.
When doing so the total number of electrons plus positrons
is changed. On the other hand the total charge remains conserved
in the presence of interactions. It is therefore the
quantity of interest.

The components of $\pc (x)$ are real numbers.
Indeed, using $(\gamma^\mu)^\dagger\gamma^0=\gamma^0\gamma^\mu$ 
one verifies that
\be
\overline{\pc ^\mu(x)}&=&
\overline{\Tr \gamma^\mu \Ga(x,x)}\cr
&=&\Tr{\Ga}^\dagger(x,x)\gamma^0\gamma^\mu\gamma^0\cr
&=&\sum_{r,r',r''}
\overline{
\langle\psi_{\kk}|[\hat\psi_{r'',\kk}(x)]^\dagger\hat\psi_{r',\kk'}(x)\psi_{\kk'}\rangle}
\gamma^0_{r,r''}[\gamma^0\gamma^\mu\gamma^0]_{r',r}
\cr
&=&\sum_{r',r''}
\int\upd\kk \,\frac{\lu^{5/2}c^{1/2}}{\sqrt{(2\pi)^32\omega(\kk)}}
\int\upd\kk' \,\frac{\lu^{5/2}c^{1/2}}{\sqrt{(2\pi)^32\omega(\kk')}}
\langle\psi_{\kk'}|[\hat\psi_{r',\kk'}(x)]^\dagger\hat\psi_{r'',\kk}(x)\psi_{\kk}\rangle
[\gamma^0\gamma^\mu]_{r',r''}
\cr
&=&\sum_{r,r''}\Ga_{r'',r}(x,x)\gamma^\mu_{r,r''}\cr
&=&\Tr \Ga(x,x)\gamma^\mu\cr
&=&\pc ^\mu(x).
\ee

The current operator $\hat\Pc (x)$ corresponding to the classical current $\pc (x)$
equals
\be
\hat\Pc^\mu_{\kk,\kk'}(x)&=&\sum_{r,r'}\psicha_{r,\kk}(x)\gamma^\mu_{r,r'}\hat\psi_{r',\kk'}(x).
\ee
Indeed, one can write
\be
\pc^\mu(x)
&=&
\int\upd\kk \,\frac{\lu^{5/2}c^{1/2}}{\sqrt{(2\pi)^32\omega(\kk)}}
\int\upd\kk' \,\frac{\lu^{5/2}c^{1/2}}{\sqrt{(2\pi)^32\omega(\kk')}}
\langle\psi_{\kk}|\hat\Pc^\mu_{\kk,\kk'}(x)\psi_{\kk'}\rangle.
\ee
The zeroth component is a density operator. It simplifies to
\be
\hat\Pc^0_{\kk,\kk'}(x)&=&\sum_r\left(\hat\psi_{r,\kk}(x)\right)^\dagger\hat\psi_{r,\kk'}(x).
\label{fermion:pdo}
\ee
This is a positive operator, in the sense that for any wave function $\psi$ one has
$\pc^0(x)\ge 0$.
The integral over space is a constant of the motion.
However, it turns out that the integral diverges for wave functions of interest.

For further use let us show that
\be
\left(\hat\Pc^\mu_{\kk,\kk'}(x)\right)^\dagger&=&\hat\Pc^\mu_{\kk',\kk}(x).
\label{electron:sa}
\ee
One calculates, using $\gamma^0[\gamma^\mu]^\dagger=\gamma^\mu\gamma^0$,
\be
\left(\hat\Pc^\mu_{\kk,\kk'}(x)\right)^\dagger
&=&
\sum_{r,r'}\overline{\gamma^\mu_{r,r'}}
\left(\psicha_{r',\kk'}(x)\hat\psi_{r,\kk}(x)
\right)^\dagger\cr
&=&\sum_{r,r'}[\gamma^\mu]^\dagger_{r',r}
\left(\hat\psi_{r,\kk}(x)\right)^\dagger
\sum_{r''}\overline{\gamma^0_{r'',r'}}\hat\psi_{r'',\kk'}(x)\cr
&=&\sum_{r,r',r''}\gamma^0_{r',r}
\left(\hat\psi_{r,\kk}(x)\right)^\dagger
\gamma^\mu_{r'',r'}\hat\psi_{r'',\kk'}(x)\cr
&=&\sum_{r',r''}\psicha_{r',\kk}(x)
\gamma^\mu_{r'',r'}\hat\psi_{r'',\kk'}(x)\cr
&=&\hat\Pc^\mu_{\kk',\kk}(x).
\ee
This proves (\ref {electron:sa}).

\subsection{The electric current}

The electric current operator is defined by
\be
[\hat j^\mu(x)\psi]_{\kk}
&=&\frac {\uc c}{(2\pi)^3}\int\upd\kk'\, \hat J^\mu_{\kk,\kk'}(x)\psi_{\kk'}.
\label{electron:jdef}
\ee
with
\be
\hat J^\mu_{\kk,\kk'}(x)&=&\frac 12\left(
\hat\Pc^\mu_{\kk,\kk'}(x)-\hat C\hat\Pc^\mu_{\kk,\kk'}(x)\hat C^{-1}\right).
\ee
Here, $\uc $ is a unit of charge.
Because $\hat\pc$ satisfies the continuity equation also $\hat j$ does.
It satisfies the reality condition
\be
\overline{\int\upd\kk\,\langle\phi|\hat j\psi\rangle_\kk}
&=&\int\upd\kk\,\langle\psi|\hat j\phi\rangle_\kk.
\ee

\def\exb{\mbox{\tiny{diag}}}
\def\ca{\mbox{\tiny{off}}}

It is shown in the Appendix \ref {appendix:explicitcurrent} that
\be
\hat J^\mu_{\kk,\kk'}(x)
&=&\frac {1}2
\sum_{r,r'}\gamma^\mu_{r,r'}\psicha_{r,\kk}(x)\hat\psi_{r',\kk'}(x)
-\frac {1}2\sum_{r,r'}\gamma^\mu_{r',r}\hat\psi_{r,\kk}(x)\psicha_{r',\kk'}(x).
\label{appE:Jexpr}
\ee
This is a well-known expression for the Dirac current, adapted to the present context.

The current operator can be split into two contributions,
$\hat J^{\exb,\mu}(x)$, which commutes with the number operator $\hat N$, and
the off-diagonal part $\hat J^{\ca,\mu}(x)$.
Using the definitions of $\hat\psi$ and $\psicha$ the expression
(\ref {appE:Jexpr}) can be further evaluated.
It is shown in the same appendix that 
\be
\hat J^\mu(x)
&=&
\hat J^{\exb,\mu}(x)+\hat J^{\ca,\mu}(x)
\ee
with
\be
\hat J^{\exb,\mu}_{\kk,\kk'}(x)
&=&
\frac 12\sum_{s,t=1,2}
\langle u^{(s)}(\kk)|\gamma^0\gamma^\mu u^{(t)}(\kk')\rangle
\hat\phi_{s,\kk}^{(-)}(x)\hat\phi_{t,\kk'}^{(+)}(x)\cr
& &
-\frac 12
\sum_{s,t=3,4}
\langle v^{(s)}(\kk')|\gamma^0\gamma^\mu v^{(t)}(\kk)\rangle
\hat\phi_{t,\kk}^{(-)}(x)\hat\phi_{s,\kk'}^{(+)}(x)\cr
& &
+(\kk\leftrightarrow\kk')
\label{el:cura}
\ee
and
\be
\hat J^{\ca,\mu}_{\kk,\kk'}(x)
&=&
\frac {1}2\sum_{s=1,2}\sum_{t=3,4}
\langle u^{(s)}(\kk)|\gamma^0\gamma^\mu v^{(t)}(\kk')\rangle\,
\hat\phi_{s,\kk}^{(-)}(x)\hat\phi_{t,\kk'}^{(-)}(x)\cr
& &
+\frac {1}2\sum_{s=3,4}\sum_{t=1,2}
\langle v^{(s)}(\kk)|\gamma^0\gamma^\mu u^{(t)}(\kk')\rangle\,
\hat\phi_{s,\kk}^{(+)}(x)\hat\phi_{t,\kk'}^{(+)}(x)\cr
& &
+(\kk\leftrightarrow\kk').
\label{el:curb}
\ee
Note that these are normal-ordered expressions.

One verifies that each of the two current operators $\hat J^{\exb,\mu}(x)$ and $\hat J^{\ca,\mu}(x)$ satisfies
the continuity equation
\be
i\partial_\mu \hat J^{\exb,\mu}_{\kk,\kk'}(x)&=&0,
\label{current:diag:cont}
\ee
respectively
\be
i\partial_\mu \hat J^{\ca,\mu}_{\kk,\kk'}(x)&=&0
\label{current:off:cont}
\ee
See the Appendix \ref {app:cont}.
In addition they satisfy
\be
k_\mu \hat J^{\exb,\mu}_{\kk,\kk'}(x)
&=&k'_\mu \hat J^{\exb,\mu}_{\kk,\kk'}(x)
\label{current:diag:sym}
\ee
and
\be
k_\mu \hat J^{\ca,\mu}_{\kk,\kk'}(x)
&=&-k'_\mu \hat J^{\ca,\mu}_{\kk,\kk'}(x).
\label{current:off:sym}
\ee

Let us now calculate the total charge $\hat Q$, which is defined by
\be
\hat Q&=&\frac 1c\int\upd \xx \,\hat j^0(x).
\label{electron:totcharge}
\ee
Using the orthogonality relations (\ref {electron:ortho}) one finds
\be
\int\upd\xx\,\hat J^0_{\kk,\kk'}(x)
&=&
(2\pi)^3\delta(\kk-\kk')\left(
\hat N^{(1)}+\hat N^{(2)}-\hat N^{(3)}-\hat N^{(4)}\right).
\ee
Hence, for any wave function $\psi_\kk$ is
\be
(\hat Q\psi)_\kk
&=&
\frac 1c\int\upd \xx \,(\hat j^0(x)\psi)_\kk\cr
&=&
\frac {\uc c}{(2\pi)^3}\int\upd\kk'\, 
\frac 1c\int\upd \xx \,\hat J^\mu_{\kk,\kk'}(x)\psi_{\kk'}\cr
&=&
\uc \int\upd\kk'\, 
\delta(\kk-\kk')\left(
\hat N^{(1)}+\hat N^{(2)}-\hat N^{(3)}-\hat N^{(4)}\right)\psi_{\kk'}\cr
&=&\uc 
\left(
\hat N^{(1)}+\hat N^{(2)}-\hat N^{(3)}-\hat N^{(4)}\right)\psi_{\kk}.
\ee
One concludes that
\be
\hat Q&=&\uc\left(\hat N^{(1)}+\hat N^{(2)}-\hat N^{(3)}-\hat N^{(4)}\right).
\label{electron:totchargeresult}
\ee
The obvious interpretation is that the components 1 and 2 of the field describe an electron
with charge $\uc $, and that components 3 and 4 describe a positron with charge $-\uc $.

Note that
\be
[\hat\phi^{(+)}_{s,\kk},\hat Q]_{_-}&=&\uc \hat\phi^{(+)}_{s,\kk},
\qquad s=1,2,\cr
[\hat\phi^{(-)}_{s,\kk},\hat Q]_{_-}&=&\uc \hat\phi^{(-)}_{s,\kk},
\qquad s=3,4.
\ee
This implies
\be
\left[\hat\psi_{r,\kk}(x),\hat Q\right]_-&=&q\hat\psi_{r,\kk}(x),
\qquad r=1,2,3,4.
\ee
By taking the Hermitean conjugate one obtains
\be
\left[\hat\psi^\dagger_{r,\kk}(x),\hat Q\right]_-&=&-q\hat\psi^\dagger_{r,\kk}(x),
\qquad r=1,2,3,4.
\ee
From (\ref {appE:Jexpr}) then follows that
\be
\left[\hat J^\mu_{\kk,\kk'}(x),\hat Q\right]_-&=&0.
\ee

\subsection{Example of a polarized electron field}
\label{electron:example}

Assume a wave function of the form (\ref {fermion:wf}).
Adapted to the present context it looks like
\be
\psi_\kk&=&e^{i\chi(\kk)}\sqrt{1-\rho(\kk)}|\emptyset\rangle
+e^{i\xi(\kk)}\sqrt{\rho(\kk)}|\{1\}\rangle.
\ee
It describes an electron field polarized with spin up.

Take 
\be
\rho(\kk)=\frac{1}{\cosh^2(a|\kk|)}
\quad\mbox{and}\quad\xi(\kk)=\chi(\kk)=0.
\label{electron:exrho}
\ee
The classical Dirac fields equal
\be
\phi_r(x)
&=&
\int\upd\kk\,\frac{\lu^{5/2}c^{1/2}}{\sqrt{(2\pi)^32\omega(\kk)}}
\langle \psi_\kk|\hat\psi_{r,\kk}(x)\psi_\kk\rangle\cr
&=&
\int\upd\kk\,\frac{\lu^{5/2}c^{1/2}}{\sqrt{(2\pi)^32\omega(\kk)}}
\sum_{s=1,2}u^{(s)}_r(\kk)\langle \psi_\kk|\hat\phi^{(+)}_{s,\kk}(x)\psi_\kk\rangle\cr
&=&
\int\upd\kk\,\frac{\lu^{5/2}c^{1/2}}{\sqrt{(2\pi)^32\omega(\kk)}}
u^{(1)}_r(\kk)\sqrt{\rho(\kk)(1-\rho(\kk))}
\langle\emptyset|\hat\phi^{(+)}_{1,\kk}(x)|\{1\}\rangle\cr
&=&
\int\upd\kk\,\frac{\lu^{5/2}}{\sqrt{(2\pi)^32k_0}}
u^{(1)}_r(\kk)\frac{\tanh(a|\kk|)}{\cosh(a|\kk|)}e^{-ik_0x^0}e^{i\kk\cdot\xx}.
\ee
The classical energy is (see (\ref {fermion:classenerg}))
\be
\Ecl
&=&\lu^3\int\upd\kk\,\hbar\omega(\kk)\rho(\kk)(1-\rho(\kk))\cr
&=&\lu^3\int\upd\kk\,\hbar\omega(\kk)\frac{\tanh^2(a|\kk|)}{\cosh^2(a|\kk|)}\cr
&=&\frac{4\pi \hbar c\lu^3}{a^4}\int_0^\infty\upd r\,r^2\sqrt{\kappa^2+\left(\frac ra\right)^2} \frac{\tanh^2(r)}{\cosh^2(r)},
\ee
which is less than the quantum energy (see (\ref {fermion:quantenerg}))
\be
E&=&\langle\hat H\rangle\cr
&=&\lu^3\int\upd\kk\,\hbar\omega(\kk)\rho(\kk)\cr
&=&\lu^3\int\upd\kk\,\hbar\omega(\kk)\frac{1}{\cosh^2(a|\kk|)}\cr
&=&\frac{4\pi \hbar c\lu^3}{a^3}\int_0^\infty\upd r\,r^2\sqrt{\kappa^2+\left(\frac ra\right)^2} \frac{1}{\cosh^2(r)}.
\label{el:ex:E}
\ee
A short calculation yields a total charge given by
\be
Q
&=&\lu^3\int\upd\kk\,\langle\psi_\kk|\hat Q\psi_\kk\rangle\cr
&=&\uc \lu^3\int\upd\kk\,\rho(\kk)\cr
&=&\frac{q 4\pi\lu^3}{a^3}\int_0^\infty\upd r\,r^2\frac 1{\cosh^2(r)}\cr
&=&\frac{q \pi^3\lu^3}{3a^3}.
\label{el:ex:q}
\ee
In the limit of large $a$ one obtains from the combination of (\ref {el:ex:E})
and (\ref {el:ex:q})
\be
E&\simeq&mc^2\frac Qq.
\ee
If the field contains exactly one electron then the total charge $Q$ equals the elementary charge $q$
and the total energy in the long-wavelength limit is the rest mass energy of a single electron.

The above discussion does not depend on the choice of the length scale $\lu$, as it should be.
However, the electron field has an intrinsic length scale
$\mstar^{-1}=\hbar/mc\simeq 4\times 10^{-13}$m.
It is therefore obvious that $\lu$ should equal $1/\mstar$, up to some numerical factor.


\section{Summary}

In the present formalism free electromagnetic waves are described by providing a 
two-dimensional quantum harmonic oscillator for each value of the wave vector 
$\kk$. By analogy a 16-dimensional Hilbert space is introduced for each given 
wave vector $\kk$ to describe the spin and charge degrees of freedom of the 
electron field. The space time dimensions increase from 4 to 6 to
accommodate the electromagnetic waves. No extra dimensions are needed for the
electron field. It is more appropriate to see the electron field as a
deformation of the vacuum through which the electromagnetic waves propagate.

Field operators $\hat\psi_r(x)$ satisfying the free Dirac equation have been constructed
in Section \ref {sect:electron}. Electric current operators $\hat j_\mu(x)$
have been introduced as pair correlation functions. A short calculation brings them
into the familiar form of Dirac currents. A small example shows that
the relevant quantities are well-defined. In particular, the spinor of classical fields
obtained by taking the quantum expectations of the field operators satisfies
the original Dirac equation \cite {DPAM28}.

The difficult part of the theory of quantum electrodynamics is of course
the photon-electron interaction. It can be split up in two parts: the interaction
of electrons with free photons and the interaction of the electron field
with itself via the electromagnetic field it produces and its interaction with the charge of
the field. The latter part is treated in the next of this series of papers.

\appendix 
\section*{Appendices}

\section{Polarization of electron waves}
\label{appendix:polelectron:section}

Let us first consider solutions of the equation
\be
\gamma^\mu k_\mu u^{(s)}_{\kk}&=&\mstar u^{(s)}_{\kk}.
\label{app:polectron:ueq}
\ee
When $\kk=0$ the equation reduces to
\be
\gamma^0 u^{(s)}_{0}=u^{(s)}_{0}.
\ee
This has indeed two independent solutions $u^{(1)}_0$ and $u^{(1)}_0$.
Assuming the standard representation of the gamma matrices one can choose
$(1,0,0,0)^{\rm T}$ and $(0,1,0,0)^{\rm T}$.

Next, choose $u^{(s)}_{r,\kk}$ of the form
\be
u^{(s)}_{r}(\kk)
&=&\frac{1}{\sqrt{2k_0}}\frac{1}{\sqrt{k_0+\mstar}}\left[
\sum_{r'}k_\nu\gamma^\nu_{r,r'}u^{(s)}_{r',0} +\mstar u^{(s)}_{r,0}\right].
\label{app:polelectron:usol}
\ee
Then one finds, using $\gamma^\mu\gamma^\nu+\gamma^\nu\gamma^\mu=2g^{\mu,\nu}$,
\be
k_\mu\gamma^\mu  u^{(s)}_{\kk}
&=&\frac{1}{\sqrt{2k_0}}\frac{1}{\sqrt{k_0+\mstar}}\left[
k_\mu k_\nu \gamma^\mu \gamma^\nu u^{(s)}_{0}+\mstar  k_\mu\gamma^\mu u^{(s)}_{0}\right]\cr
&=&\frac{1}{\sqrt{2k_0}}\frac{1}{\sqrt{k_0+\mstar}}\left[
k_\mu k^\mu u^{(s)}_{0}+\mstar  k_\mu\gamma^\mu u^{(s)}_{0}\right]\cr
&=&\frac{1}{\sqrt{2k_0}}\frac{1}{\sqrt{k_0+\mstar}}\left[
\mstar^2 u^{(s)}_{0}+\mstar  k_\mu\gamma^\mu u^{(s)}_{0}\right]\cr
&=&\mstar u^{(s)}_{\kk}.
\ee
One concludes that $u^{(s)}_{r,\kk}$, defined by (\ref {app:polelectron:usol}),
solves the equation (\ref {app:polectron:ueq}).

Next one verifies that
\be
\sum_r\overline{u^{(s)}_r}(\kk)u^{(s')}_r(\kk)
&=&
\frac{1}{2k_0}\frac{1}{k_0+\mstar}
\left\langle u_0^{(s)}\big|
(k_\nu(\gamma^\nu)^\dagger+\mstar)(k_\sigma\gamma^\sigma+\mstar)u_0^{(s')}\right\rangle\cr
&=&
\frac{1}{2k_0}\frac{1}{k_0+\mstar}
\bigg[
\left\langle u_0^{(s)}\big|
k_\nu(\gamma^\nu)^\dagger k_\sigma\gamma^\sigma u_0^{(s')}\right\rangle\cr
& &
+\mstar \left\langle u_0^{(s)}\big|
k_\sigma\gamma^\sigma u_0^{(s')}\right\rangle
+\mstar \left\langle u_0^{(s)}\big|
k_\nu(\gamma^\nu)^\dagger u_0^{(s')}\right\rangle
+\mstar^2\bigg].
\ee
Now use that $(\gamma^\nu)^\dagger\gamma^0=\gamma^0\gamma^\nu$
and that $\gamma^0 u^{(s)}_{0}=u^{(s)}_{0}$
to obtain
\be
k_\nu(\gamma^\nu)^\dagger k_\sigma\gamma^\sigma u^{(s)}_{0}
&=&
\gamma^0k_\nu\gamma^\nu \gamma^0 k_\sigma\gamma^\sigma u^{(s)}_{0}\cr
&=&\left(k_0-\sum_\alpha k_\alpha\gamma^\alpha\right)
\left(k_0+\sum_\beta k_\beta\gamma^\beta\right)u^{(s)}_{0}\cr
&=&(k_0^2+|\kk|^2)u^{(s)}_{0}.
\label{app:polectron:id}
\ee
Because also
\be
\left\langle u_0^{(s)}\big|
k_\sigma\gamma^\sigma u_0^{(s')}\right\rangle
&=&k_0
\ee
there follows
\be
\sum_r\overline{u^{(s)}_r}(\kk)u^{(s')}_r(\kk)
&=&\delta_{s,s'}.
\ee

Next consider the equation
\be
\gamma^\mu k_\mu v^{(s)}_{\kk}&=&-\mstar v^{(s)}_{\kk}.
\label{app:polectron:veq}
\ee
At $\kk=0$ it becomes $\gamma^0v^{(s)}_{0}=- v^{(s)}_{0}$.
The matrix $\gamma^0$ has two eigenvectors with eigenvalue $-1$.
We choose
\be
v^{(3)}_{0}=(0,0,1,0)^{\rm T}
\quad\mbox{and}\quad
v^{(4)}_{0}=(0,0,0,1)^{\rm T}.
\ee
Next one verifies that
\be
v^{(s)}_{r}(\kk)
&=&\frac{1}{\sqrt{2k_0}}\frac{1}{\sqrt{k_0+\mstar}}\left[
-\sum_{r'}k_\nu\gamma^\nu_{r,r'}v^{(s)}_{r',0} +\mstar v^{(s)}_{r,0}\right]
\label{app:polelectron:vsol}
\ee
is a solution of (\ref {app:polectron:veq}).
The normalization
\be
\sum_r\overline{v^{(s)}_r}(\kk)v^{(s')}_r(\kk)&=&\delta_{s,s'}
\ee
is proved in the same way as for the u-vectors.

Finally let us calculate
\be
\sum_r\overline{u^{(s)}_r}(\kk)v^{(s')}_r(-\kk)
&=&\frac{1}{2k_0}\frac{1}{k_0+\mstar}
\left\langle u_0^{(s)}\big|
(k_0\gamma^0-\sum_\alpha k_\alpha\gamma^\alpha+\mstar)
(-k_0\gamma^0+\sum_\beta k_\beta\gamma^\beta+\mstar)v_0^{(s')}\right\rangle\cr
&=&-\frac{1}{2k_0}\frac{1}{k_0+\mstar}
\left\langle u_0^{(s)}\big|
(k_0+\mstar)^2-\left(\sum_\alpha k_\alpha\gamma^\alpha\right)^2v_0^{(s')}\right\rangle\cr
&=&
-\frac{1}{2k_0}\frac{1}{k_0+\mstar}
\left((k_0+\mstar)^2+|\kk|^2\right)
\langle u_0^{(s)}v_0^{(s')}\rangle\cr
&=&0.
\ee
This ends the verification of the orthogonality relations.

The inverse relations (\ref {electron:inv+}, \ref {electron:inv-}) follow easily.
Using the orthogonality relations one obtains
\be
\sum_r\overline{u_r^{(s)}(-\kk)}\hat\psi_{r,\kk}(x)
&=&\sum_{t=1,2}\langle u^{(s)}(-\kk)|u^{(t)}(\kk)\rangle\hat\phi^{(+)}_{t,\kk}(x).
\label{app:tempinv1}
\ee
Now use the definitions to evaluate
\be
\langle u^{(s)}(-\kk)|u^{(t)}(\kk)\rangle
&=&
\frac{1}{2k_0(k_0+\mstar)}
\langle(k_0+\mstar-\sum_\alpha k_\alpha\gamma^\alpha)u_0^{(s)}|
(k_0+\mstar+\sum_\alpha k_\alpha\gamma^\alpha)u_0^{(t)}\rangle\cr
&=&
\frac{1}{2k_0(k_0+\mstar)}
\langle u_0^{(s)}|
(k_0+\mstar+\sum_\alpha k_\alpha\gamma^\alpha)^2u_0^{(t)}\rangle\cr
&=&\frac{1}{2k_0(k_0+\mstar)}
\langle u_0^{(s)}|
(k_0+\mstar)^2
+\sum_\alpha k^2_\alpha(\gamma^\alpha)^2u_0^{(t)}\rangle\cr
&=&\frac{(k_0+\mstar)^2-|\kk|^2}{2k_0(k_0+\mstar)}\delta_{s,t}\cr
&=&\frac{\mstar}{k_0}\delta_{s,t}.
\ee
Hence (\ref {app:tempinv1}) implies (\ref {electron:inv+}).
The derivation of (\ref {electron:inv-}) is similar.

\section{Useful relations}
\label{app:useful}

A useful relation is
\be
\gamma^0 u^{(s)}(\kk)
&=& \gamma^0 \frac{1}{\sqrt{2k_0}}\frac{1}{\sqrt{k_0+\mstar}}\left[
k_0\gamma^0+\sum_{\alpha=1}^3k_\alpha\gamma^\alpha +\mstar \right]u^{(s)}_{0}\cr
&=&\frac{1}{\sqrt{2k_0}}\frac{1}{\sqrt{k_0+\mstar}}\left[
k_0-\sum_{\alpha=1}^3k_\alpha\gamma^\alpha +\mstar \right]u^{(s)}_{0}\cr
&=&u^{(s)}(-\kk).
\label{app:minusk}
\ee
Similarly is $\gamma^0 v^{(s)}(\kk)=-v^{(s)}(-\kk)$.
Both relations are used in
\be
\langle v^{(4)}(\kk)|\gamma^0\gamma^\mu v^{(4)}(\kk')\rangle
&=&
\langle C\overline{u^{(1)}(-\kk)}|\gamma^0\gamma^\mu C\overline{u^{(1)}(-\kk')}\rangle\cr
&=&
\langle \overline{u^{(1)}(-\kk)}|C^\dagger\gamma^0\gamma^\mu C\overline{u^{(1)}(-\kk')}\rangle\cr
&=&
\langle \overline{u^{(1)}(-\kk)}|(\gamma^\mu\gamma^0)^{\rm T}\overline{u^{(1)}(-\kk')}\rangle\cr
&=&
\langle u^{(1)}(-\kk')|\gamma^\mu\gamma^0u^{(1)}(-\kk)\rangle\cr
&=&
\langle  u^{(1)}(\kk')|\gamma^0 \gamma^\mu u^{(1)}(\kk)\rangle.
\label{appC:8relfirst}
\ee
Similarly is
\be
\langle v^{(4)}(\kk)|\gamma^0\gamma^\mu v^{(3)}(\kk')\rangle
&=&
\langle  u^{(2)}(\kk')|\gamma^0 \gamma^\mu u^{(1)}(\kk)\rangle;\\
\langle v^{(3)}(\kk)|\gamma^0\gamma^\mu v^{(3)}(\kk')\rangle
&=&
\langle  u^{(2)}(\kk')|\gamma^0 \gamma^\mu u^{(2)}(\kk)\rangle.
\label{appC:8relsec}
\ee
One has also
\be
\langle u^{(1)}(\kk)|\gamma^0\gamma^\mu v^{(4)}(\kk')\rangle
&=&
\langle u^{(1)}(\kk)|\gamma^0\gamma^\mu C\overline{u^{(1)}(-\kk')}\rangle\cr
&=&
\langle u^{(1)}(\kk)|C(\gamma^\mu\gamma^0)^{\rm T}\overline{u^{(1)}(-\kk')}\rangle\cr
&=&
-\langle u^{(1)}(-\kk')|\gamma^\mu\gamma^0 C\overline{u^{(1)}(\kk)}\rangle\cr
&=&
-\langle u^{(1)}(-\kk')|\gamma^\mu\gamma^0 v^{(4)}(-\kk)\rangle\cr
&=&
\langle u^{(1)}(\kk')|\gamma^0\gamma^\mu v^{(4)}(\kk)\rangle.
\label{appC:8relthird}
\ee
Similarly is
\be
\langle u^{(2)}(\kk)|\gamma^0\gamma^\mu v^{(4)}(\kk')\rangle
&=&
\langle u^{(1)}(\kk')|\gamma^0\gamma^\mu v^{(3)}(\kk)\rangle;\\
\langle u^{(2)}(\kk)|\gamma^0\gamma^\mu v^{(3)}(\kk')\rangle
&=&
\langle u^{(2)}(\kk')|\gamma^0\gamma^\mu v^{(3)}(\kk)\rangle.
\label{appC:8rellast}
\ee

Next we calculate
\be
& &\langle u^{(s)}(\kk)|\gamma^0\gamma^\mu u^{(s')}(\kk)\rangle\cr
&=&\frac{1}{2k_0(k_0+\mstar)}
\langle (k_\nu\gamma^\nu+\mstar)u^{(s)}_0|
\gamma^0\gamma^\mu(k_\tau\gamma^\tau+\mstar)u^{(s')}_0\rangle\cr
&=&\frac{1}{2k_0(k_0+\mstar)}
\langle u^{(s)}_0|
(k_\nu\gamma^\nu+\mstar)\gamma^\mu(k_\tau\gamma^\tau+\mstar)u^{(s')}_0\rangle\cr
&=&\frac{1}{2k_0(k_0+\mstar)}
\langle u^{(s)}_0|
[\gamma^\mu(-k_\nu\gamma^\nu+\mstar)+2k^\mu](k_\tau\gamma^\tau+\mstar)u^{(s')}_0\rangle\cr
&=&\frac{k^\mu}{k_0(k_0+\mstar)}\langle u^{(s)}_0|
(k_\tau\gamma^\tau+\mstar)u^{(s')}_0\rangle
+\frac{1}{2k_0(k_0+\mstar)}
\langle u^{(s)}_0|
\gamma^\mu[-k_\nu \gamma^\nu k_\tau\gamma^\tau+\mstar^2 ]u^{(s')}_0\rangle\cr
&=&\frac{k^\mu}{k_0}\delta_{s,s'}.
\label{appC:ueq}
\ee

Similarly  is
\be
& &\langle v^{(s)}(\kk)|\gamma^0\gamma^\mu v^{(s')}(\kk)\rangle\cr
&=&\frac{1}{2k_0(k_0+\mstar)}
\langle (-k_\nu\gamma^\nu+\mstar)v^{(s)}_0|
\gamma^0\gamma^\mu(-k_\tau\gamma^\tau+\mstar)v^{(s')}_0\rangle\cr
&=&\frac{1}{2k_0(k_0+\mstar)}
\langle v^{(s)}_0|
(-k_\nu\gamma^\nu+\mstar)^\dagger
\gamma^0\gamma^\mu(-k_\tau\gamma^\tau+\mstar)v^{(s')}_0\rangle\cr
&=&-\frac{1}{2k_0(k_0+\mstar)}
\langle v^{(s)}_0|
(-k_\nu\gamma^\nu+\mstar)\gamma^\mu(-k_\tau\gamma^\tau+\mstar)v^{(s')}_0\rangle\cr
&=&-\frac{1}{2k_0(k_0+\mstar)}
\langle v^{(s)}_0|
[\gamma^\mu(k_\nu\gamma^\nu+\mstar)-2k^\mu](-k_\tau\gamma^\tau+\mstar)v^{(s')}_0\rangle\cr
&=&\frac{k^\mu}{k_0(k_0+\mstar)}\langle v^{(s)}_0|
(-k_\tau\gamma^\tau+\mstar)v^{(s')}_0\rangle
-\frac{1}{2k_0(k_0+\mstar)}
\langle v^{(s)}_0|
\gamma^\mu[-k_\nu \gamma^\nu k_\tau\gamma^\tau+\mstar^2 ]v^{(s')}_0\rangle\cr
&=&\frac{k^\mu}{k_0}\delta_{s,s'}.
\label{appC:veq}
\ee
This finishes the proof of (\ref {electron:uid}, \ref {electron:vid}).

\section{The current operators}
\label {appendix:explicitcurrent}

Here explicit expressions for the current operators $\hat J^\mu_{\kk,\kk'}(x)$ are calculated.

From the definition follows
\be
\hat J^\mu_{\kk,\kk'}(x)
&=&\frac {1}2\left(
\hat\Pc^\mu_{\kk,\kk'}(x)-\hat C\hat\Pc^\mu_{\kk,\kk'}(x)\hat C^{-1}\right)\cr
&=&\frac {1}2
\sum_{j,j'}\gamma^\mu_{j,j'}\left(\psicha_{j,\kk}(x)\hat\psi_{j',\kk'}(x)
-\hat C\psicha_{j,\kk}(x)\hat\psi_{j',\kk'}(x)\hat C^{-1}\right)\cr
&=&\frac {1}2\left(
\sum_{j,j'}\gamma^\mu_{j,j'}\psicha_{j,\kk}(x)\hat\psi_{j',\kk'}(x)
+\sum_{j,j'}\gamma^\mu_{',j'}\sum_{r,r'}C_{j,r}\hat\psi_{r,\kk}(x)C_{j',r'}\psicha_{r',\kk'}(x)\right)\cr
&=&\frac {1}2\left(
\sum_{j,j'}\gamma^\mu_{j,j'}\psicha_{j,\kk}(x)\hat\psi_{j',\kk'}(x)
+\sum_{j,j'}(C^{\rm T}\gamma^\mu C)_{j,j'}\hat\psi_{j,\kk}(x)\psicha_{j',\kk'}(x)\right)\cr
& &
\ee
Use that $C^{\rm T}\gamma^\mu C=-(\gamma^\mu)^{\rm T}$ to obtain 
\be
\hat J^\mu_{\kk,\kk'}(x)
&=&\frac {1}2
\sum_{j,j'}\gamma^\mu_{j,j'}\psicha_{j,\kk}(x)\hat\psi_{j',\kk'}(x)
-\frac {1}2\sum_{j,j'}\gamma^\mu_{j',j}\hat\psi_{j,\kk}(x)\psicha_{j',\kk'}(x).
\ee
This is (\ref {appE:Jexpr}).

Use the definition of the adjoint operators to obtain
\be
\hat J^\mu_{\kk,\kk'}(x)
&=&\frac {1}2
\sum_{j,j'}(\gamma^0\gamma^\mu)_{j,j'}\hat\psi^\dagger_{j,\kk}(x)\hat\psi_{j',\kk'}(x)
-\frac {1}2\sum_{j,j'}(\gamma^0\gamma^\mu)_{j,j'}\hat\psi_{j',\kk}(x)\hat\psi^\dagger_{j,\kk'}(x).
\label{appE:temp}
\ee
Now use the definition of the $\hat\psi_j$. The two terms of (\ref {appE:temp}) are evaluated
separately. One has
\be
\frac {1}2
\sum_{j,j'}(\gamma^0\gamma^\mu)_{j,j'}\hat\psi^\dagger_{j,\kk}(x)\hat\psi_{j',\kk'}(x)
&=&
\frac {1}2
\sum_{j,j'}(\gamma^0\gamma^\mu)_{j,j'}
\bigg(
\sum_{s=1,2}\overline{u_j^{(s)}(\kk)}\hat\phi_{s,\kk}^{(-)}(x)
+\sum_{s=3,4}\overline{v_j^{(s)}(\kk)}\hat\phi_{s,\kk}^{(+)}(x)\bigg)\cr
& &\qquad
\times\bigg(
\sum_{t=1,2}u_{j'}^{(t)}(\kk')\hat\phi_{t,\kk'}^{(+)}(x)
+\sum_{t=3,4}v_{j'}^{(t)}(\kk')\hat\phi_{t,\kk'}^{(-)}(x)\bigg)\cr
&=&
\frac {1}2\sum_{s,t=1,2}
\langle u^{(s)}(\kk)|\gamma^0\gamma^\mu u^{(t)}(\kk')\rangle\,
\hat\phi_{s,\kk}^{(-)}(x)\hat\phi_{t,\kk'}^{(+)}(x)\cr
& &+
\frac {1}2\sum_{s=1,2}\sum_{t=3,4}
\langle u^{(s)}(\kk)|\gamma^0\gamma^\mu v^{(t)}(\kk')\rangle\,
\hat\phi_{s,\kk}^{(-)}(x)\hat\phi_{t,\kk'}^{(-)}(x)\cr
& &+
\frac {1}2\sum_{s=3,4}\sum_{t=1,2}
\langle v^{(s)}(\kk)|\gamma^0\gamma^\mu u^{(t)}(\kk')\rangle\,
\hat\phi_{s,\kk}^{(+)}(x)\hat\phi_{t,\kk'}^{(+)}(x)\cr
& &+
\frac {1}2\sum_{s,t=3,4}
\langle v^{(s)}(\kk)|\gamma^0\gamma^\mu v^{(t)}(\kk')\rangle\,
\hat\phi_{s,\kk}^{(+)}(x)\hat\phi_{t,\kk'}^{(-)}(x)\cr
&=&
\frac {1}2\sum_{s,t=1,2}
\langle u^{(s)}(\kk)|\gamma^0\gamma^\mu u^{(t)}(\kk')\rangle\,
\hat\phi_{s,\kk}^{(-)}(x)\hat\phi_{t,\kk'}^{(+)}(x)\cr
& &+
\frac {1}2\sum_{s=1,2}\sum_{t=3,4}
\langle u^{(s)}(\kk)|\gamma^0\gamma^\mu v^{(t)}(\kk')\rangle\,
\hat\phi_{s,\kk}^{(-)}(x)\hat\phi_{t,\kk'}^{(-)}(x)\cr
& &+
\frac {1}2\sum_{s=3,4}\sum_{t=1,2}
\langle v^{(s)}(\kk)|\gamma^0\gamma^\mu u^{(t)}(\kk')\rangle\,
\hat\phi_{s,\kk}^{(+)}(x)\hat\phi_{t,\kk'}^{(+)}(x)\cr
& &-
\frac {1}2\sum_{s,t=3,4}
\langle v^{(s)}(\kk)|\gamma^0\gamma^\mu v^{(t)}(\kk')\rangle\,
\hat\phi_{t,\kk'}^{(-)}(x)\hat\phi_{s,\kk}^{(+)}(x)\cr
& &+
\frac {1}2\sum_{s=3,4}
\langle v^{(s)}(\kk)|\gamma^0\gamma^\mu v^{(s)}(\kk')\rangle\,
e^{i(k'_\nu-k_\nu)x^\nu}.
\ee
The second term becomes
\be
\frac {1}2\sum_{j,j'}(\gamma^0\gamma^\mu)_{j,j'}\hat\psi_{j',\kk}(x)\hat\psi^\dagger_{j,\kk'}(x)
&=&
\frac {1}2
\sum_{j,j'}(\gamma^0\gamma^\mu)_{j,j'}
\bigg(
\sum_{t=1,2}u_{j'}^{(t)}(\kk)\hat\phi_{t,\kk}^{(+)}(x)
+\sum_{t=3,4}v_{j'}^{(t)}(\kk)\hat\phi_{t,\kk}^{(-)}(x)\bigg)\cr
& &\times
\bigg(
\sum_{s=1,2}\overline{u_j^{(s)}(\kk')}\hat\phi_{s,\kk'}^{(-)}(x)
+\sum_{s=3,4}\overline{v_j^{(s)}(\kk')}\hat\phi_{s,\kk'}^{(+)}(x)\bigg)\cr
&=&
\frac {1}2
\sum_{s,t=1,2}
\langle u^{(s)}(\kk')|\gamma^0\gamma^\mu u^{(t)}(\kk)\rangle\,
\hat\phi_{t,\kk}^{(+)}(x)\hat\phi_{s,\kk'}^{(-)}(x)\cr
& &+
\frac {1}2
\sum_{s=1,2}\sum_{t=3,4}
\langle u^{(s)}(\kk')|\gamma^0\gamma^\mu v^{(t)}(\kk)\rangle\,
\hat\phi_{t,\kk}^{(-)}(x)\hat\phi_{s,\kk'}^{(-)}(x)\cr
& &+
\frac {1}2
\sum_{s=3,4}\sum_{t=1,2}
\langle v^{(s)}(\kk')|\gamma^0\gamma^\mu u^{(t)}(\kk)\rangle\,
\hat\phi_{t,\kk}^{(+)}(x)\hat\phi_{s,\kk'}^{(+)}(x)\cr
& &+
\frac {1}2\sum_{s,t=3,4}
\langle v^{(s)}(\kk')|\gamma^0\gamma^\mu v^{(t)}(\kk)\rangle\,
\hat\phi_{t,\kk}^{(-)}(x)\hat\phi_{s,\kk'}^{(+)}(x)\cr
&=&
-\frac {1}2
\sum_{s,t=1,2}
\langle u^{(s)}(\kk')|\gamma^0\gamma^\mu u^{(t)}(\kk)\rangle\,
\hat\phi_{s,\kk'}^{(-)}(x)\hat\phi_{t,\kk}^{(+)}(x)\cr
& &+
\frac {1}2
\sum_{s=1,2}
\langle u^{(s)}(\kk')|\gamma^0\gamma^\mu u^{(s)}(\kk)\rangle\,
e^{i(k'_\nu-k_\nu)x^\nu}\cr
& &+
\frac {1}2
\sum_{s=1,2}\sum_{t=3,4}
\langle u^{(s)}(\kk')|\gamma^0\gamma^\mu v^{(t)}(\kk)\rangle\,
\hat\phi_{t,\kk}^{(-)}(x)\hat\phi_{s,\kk'}^{(-)}(x)\cr
& &+
\frac {1}2
\sum_{s=3,4}\sum_{t=1,2}
\langle v^{(s)}(\kk')|\gamma^0\gamma^\mu u^{(t)}(\kk)\rangle\,
\hat\phi_{t,\kk}^{(+)}(x)\hat\phi_{s,\kk'}^{(+)}(x)\cr
& &+
\frac {1}2\sum_{s,t=3,4}
\langle v^{(s)}(\kk')|\gamma^0\gamma^\mu v^{(t)}(\kk)\rangle\,
\hat\phi_{t,\kk}^{(-)}(x)\hat\phi_{s,\kk'}^{(+)}(x).
\ee
Next subtract the two contributions.
It is shown in the Appendix \ref {app:useful} that
\be
\sum_{t=3,4}\langle v^{(t)}(\kk')|\gamma^0\gamma^\mu v^{(t)}(\kk)\rangle
&=&\sum_{s=1,2}\langle u^{(s)}(\kk)|\gamma^0\gamma^\mu u^{(s)}(\kk')\rangle.
\label{appC:id}
\ee
Hence the two scalar terms cancel and one obtains
\be
\hat J^\mu_{\kk,\kk'}(x)
&=&
\frac {1}2\sum_{s,t=1,2}
\langle u^{(s)}(\kk)|\gamma^0\gamma^\mu u^{(t)}(\kk')\rangle\,
\hat\phi_{s,\kk}^{(-)}(x)\hat\phi_{t,\kk'}^{(+)}(x)\cr
& &
+\frac {1}2
\sum_{s,t=1,2}
\langle u^{(s)}(\kk'|\gamma^0\gamma^\mu u^{(t)}(\kk)\rangle\,
\hat\phi_{s,\kk'}^{(-)}(x)\hat\phi_{t,\kk}^{(+)}(x)\cr
& &-
\frac {1}2\sum_{s,t=3,4}
\langle v^{(s)}(\kk)|\gamma^0\gamma^\mu v^{(t)}(\kk')\rangle\,
\hat\phi_{t,\kk'}^{(-)}(x)\hat\phi_{s,\kk}^{(+)}(x)\cr
& &
-\frac {1}2\sum_{s,t=3,4}
\langle v^{(s)}(\kk')|\gamma^0\gamma^\mu v^{(t)}(\kk)\rangle\,
\hat\phi_{t,\kk}^{(-)}(x)\hat\phi_{s,\kk'}^{(+)}(x)\cr
& &+
\frac {1}2\sum_{s=1,2}\sum_{t=3,4}
\langle u^{(s)}(\kk)|\gamma^0\gamma^\mu v^{(t)}(\kk')\rangle\,
\hat\phi_{s,\kk}^{(-)}(x)\hat\phi_{t,\kk'}^{(-)}(x)\cr
& &-
\frac {1}2
\sum_{s=1,2}\sum_{t=3,4}
\langle u^{(s)}(\kk')|\gamma^0\gamma^\mu v^{(t)}(\kk)\rangle\,
\hat\phi_{t,\kk}^{(-)}(x)\hat\phi_{s,\kk'}^{(-)}(x)\cr
& &+
\frac {1}2\sum_{s=3,4}\sum_{t=1,2}
\langle v^{(s)}(\kk)|\gamma^0\gamma^\mu u^{(t)}(\kk')\rangle\,
\hat\phi_{s,\kk}^{(+)}(x)\hat\phi_{t,\kk'}^{(+)}(x)\cr
& &-
\frac {1}2
\sum_{s=3,4}\sum_{t=1,2}
\langle v^{(s)}(\kk')|\gamma^0\gamma^\mu u^{(t)}(\kk)\rangle\,
\hat\phi_{t,\kk}^{(+)}(x)\hat\phi_{s,\kk'}^{(+)}(x).
\ee
This result can be split into two pieces (\ref {el:cura}) and (\ref {el:curb}).

\section{The continuity equation}
\label{app:cont}

Here follows the proof that $\hat J^{\exb,\mu}(x)$ satisfies the continuity equation
(\ref {current:diag:cont}) and the symmetry property 
(\ref {current:diag:sym}).
The proofs of (\ref {current:off:cont}) and of (\ref {current:off:cont})
are analoguous and are omitted.

Use the explicit expression (\ref {el:cura}) to obtain
\be
i\partial_\mu \hat J^{\exb,\mu}_{\kk,\kk'}(x)
&=&
-\frac 12\sum_{s,t=1,2}
\langle u^{(s)}(\kk)|\gamma^0\gamma^\mu u^{(t)}(\kk')\rangle
(k_\mu-k'_\mu)\hat\phi_{s,\kk}^{(-)}(x)\hat\phi_{t,\kk'}^{(+)}(x)\cr
& &
+\frac 12
\sum_{s,t=3,4}
\langle v^{(s)}(\kk')|\gamma^0\gamma^\mu v^{(t)}(\kk)\rangle
(k_\mu-k'_\mu)\hat\phi_{t,\kk}^{(-)}(x)\hat\phi_{s,\kk'}^{(+)}(x)\cr
& &
+\frac 12\sum_{s,t=1,2}
\langle u^{(s)}(\kk')|\gamma^0\gamma^\mu u^{(t)}(\kk)\rangle
(k_\mu-k'_\mu)\hat\phi_{s,\kk'}^{(-)}(x)\hat\phi_{t,\kk}^{(+)}(x)\cr
& &
-\frac 12
\sum_{s,t=3,4}
\langle v^{(s)}(\kk)|\gamma^0\gamma^\mu v^{(t)}(\kk')\rangle
(k_\mu-k'_\mu)\hat\phi_{t,\kk'}^{(-)}(x)\hat\phi_{s,\kk}^{(+)}(x).
\ee
Each of the 4 contributions can be shown to vanish by use of the definition
of the polarization vectors $u^{(s)}$ and $v^{(t)}$ as solutions of the
equations (\ref {app:polectron:ueq}, \ref {app:polectron:veq}).

Use the same relations to calculate
\be
k_\mu \hat J^{\exb,\mu}_{\kk,\kk'}(x)
&=& 
\frac 12\sum_{s,t=1,2}
\langle u^{(s)}(\kk)|\gamma^0k_\mu\gamma^\mu u^{(t)}(\kk')\rangle
\hat\phi_{s,\kk}^{(-)}(x)\hat\phi_{t,\kk'}^{(+)}(x)\cr
& &
-\frac 12
\sum_{s,t=3,4}
\langle v^{(s)}(\kk')|\gamma^0k_\mu\gamma^\mu v^{(t)}(\kk)\rangle
\hat\phi_{t,\kk}^{(-)}(x)\hat\phi_{s,\kk'}^{(+)}(x)\cr
& &
+(\kk\leftrightarrow\kk')\cr
&=&\frac 12\mstar\sum_{s,t=1,2}
\langle u^{(s)}(\kk)|\gamma^0u^{(t)}(\kk')\rangle
\hat\phi_{s,\kk}^{(-)}(x)\hat\phi_{t,\kk'}^{(+)}(x)\cr
& &
-\frac 12\mstar
\sum_{s,t=3,4}
\langle v^{(s)}(\kk')|\gamma^0 v^{(t)}(\kk)\rangle
\hat\phi_{t,\kk}^{(-)}(x)\hat\phi_{s,\kk'}^{(+)}(x)\cr
& &
+(\kk\leftrightarrow\kk').
\ee
For the part with $\kk$ and $\kk'$ exchanged one can use that
the matrix $\gamma^0 k'_\mu\gamma^\mu$ is Hermitean.
The same expression is obtained when one calculates
$k'_\mu \hat J^{\exb,\mu}_{\kk,\kk'}(x)$.
One concludes that (\ref {current:diag:sym}) holds.

\section*{}

\end{document}